\documentclass[twocolumn]{aastex7}

\usepackage{color}
\usepackage{amsmath}
\usepackage{natbib}
\definecolor{ruri}{rgb}{0,0.5,0.7} 

\hypersetup{breaklinks=true, colorlinks=true, urlcolor={blue}, citecolor={blue}, linkcolor={blue}}

\newcommand{\rev}[1]{{\color{black} #1}}

\newcommand{\swift}{\emph{Swift}}

\newcommand{\nicer}{\emph{NICER}}
\newcommand{\ninjasat}{\emph{NinjaSat}}

\newcommand{\lumcgs}{erg~s$^{-1}$}
\newcommand{\nh}{($N_{\rm H}$}

\newcommand{\TUS}{Department of Physics, Tokyo University of Science, 1-3 Kagurazaka, Shinjuku, Tokyo 162-8601, Japan}

\newcommand{\RIKENPRI}{RIKEN Pioneering Research Institute (PRI), 2-1 Hirosawa, Wako, Saitama 351-0198, Japan}
\newcommand{\RIKENRAP}{RIKEN Center for Advanced Photonics (RAP), 2-1 Hirosawa, Wako, Saitama 351-0198, Japan}
\newcommand{\RIKENNishina}{RIKEN Nishina Center, 2-1 Hirosawa, Wako, Saitama 351-0198, Japan}
\newcommand{\KyotoU}{Department of Physics, Kyoto University, Kitashirakawa Oiwake, Sakyo, Kyoto 606-8502, Japan}
\newcommand{\ChibaU}{International Center for Hadron Astrophysics, Chiba University, 1-33 Yayoi, Inage, Chiba 263-8522, Japan}

\newcommand{\NCUE}{Department of Physics, National Changhua University of Education (NCUE), Changhua 50007, Taiwan}

\newcommand{\RRI}{Raman Research Institute, Astronomy and Astrophysics, C.V. Raman Avenue, Bangalore 560080, Karnataka, India}
\newcommand{\Rikkyo}{Department of Physics, Rikkyo University, 3-34-1 Nishi Ikebukuro, Toshima-ku, Tokyo 171-8501, Japan}

\newcommand{\HU}{Graduate School of Advanced Science and Engineering, Hiroshima University, 1-3-1 Kagamiyama, Higashi-Hiroshima, Hiroshima 739-8526, Japan}

\begin{document}

\title{Unveiling the Nature of Superorbital Modulation of SMC X-1 using NinjaSat}

\author[0000-0001-8551-2002]{Chin-Ping Hu}
\email[show]{cphu0821@gm.ncue.edu.tw}
\affiliation{\NCUE}

\author{Naoyuki Ota}
\email{naoyuki.ota@riken.jp}
\affiliation{\TUS}
\affiliation{\RIKENNishina}

\author{Takuya Takahashi}
\email{1225545@ed.tus.ac.jp}
\affiliation{\TUS}
\affiliation{\RIKENNishina}

\author[0009-0008-5133-9131]{Tomoshi Takeda}
\email{tomoshi-takeda@hiroshima-u.ac.jp}
\affiliation{\HU}

\author[0000-0003-1244-3100]{Teruaki Enoto}
\email{enoto.teruaki.2w@kyoto-u.ac.jp}
\affiliation{\KyotoU}
\affiliation{\RIKENRAP}

\author[0000-0002-8801-6263]{Toru Tamagawa}
\email{tamagawa@riken.jp}
\affiliation{\RIKENPRI}
\affiliation{\RIKENNishina}
\affiliation{\TUS}

\author{Biswajit Paul}
\email{bpaul.rri@gmail.com}
\affiliation{\RRI}

\author{Sota Watanabe}
\email{sota.watanabe@a.riken.jp}
\affiliation{\TUS}
\affiliation{\RIKENNishina}

\author[0000-0002-0207-9010]{Wataru Iwakiri}
\affiliation{\ChibaU}
\email{iwakiri@hepburn.s.chiba-u.ac.jp}


\author[0000-0002-6337-7943]{Tatehiro Mihara}
\affiliation{\RIKENPRI}
\email{tmihara@riken.jp}


\author[0009-0008-3926-363X]{Amira Aoyama} 
\affiliation{\TUS}
\affiliation{\RIKENPRI}
\email{amira.aoyama@riken.jp}

\author{Satoko Iwata}
\affiliation{\TUS}
\affiliation{\RIKENNishina}
\email{satoko.iwata@riken.jp}

\author{Kaede Yamasaki}
\affiliation{\TUS}
\affiliation{\RIKENNishina}
\email{kaede.yamasaki@a.riken.jp}

\author{Takayuki Kita}
\affiliation{\ChibaU}
\email{takayukikita@hepburn.s.chiba-u.ac.jp}

\author{Soma Tsuchiya}
\affiliation{\TUS}
\email{1222075@ed.tus.ac.jp}

\author{Mayu Ichibakase}
\affiliation{\Rikkyo}
\email{mayu.ichibakase@a.riken.jp}

\collaboration{all}{(NinjaSat collaboration)}
\correspondingauthor{C.-P. Hu}

\begin{abstract}
    We report a long-term, high-cadence timing and spectral observation of the X-ray pulsar SMC X-1 using \ninjasat, a 6U CubeSat in low-Earth orbit, covering nearly a full superorbital cycle. SMC X-1 is a high-mass X-ray binary exhibiting a 0.7 s X-ray pulsar and a non-stationary superorbital modulation with periods ranging from approximately 40 to 65 days. Its peak luminosity of $1.3\times10^{39}$~\lumcgs\ makes it a local analogue of ultraluminous X-ray pulsars powered by supercritical accretion. We find that the spin-up rate during the high state remains consistent with the long-term average, with no significant correlation between spin-up rate and flux. This result indicates that the modulation is primarily geometric rather than accretion-driven. The hardness ratio and spectral shape are stable throughout the entire superorbital cycle, supporting obscuration by optically thick material or energy-independent scattering. In addition, the 2--20 keV pulse profile varies with superorbital phase, which may be explained either by variable covering fraction due to geometric obscuration, or by free precession of the neutron star. This represents the first complete measurement of spin-up rate and spectral evolution across a single superorbital cycle in SMC X-1, highlighting the scientific capability of CubeSat-based observatories. 
\end{abstract}


\section{Introduction}\label{introduction}

SMC X-1 is an accreting X-ray pulsar in a high-mass X-ray binary (HMXB) system with a Roche-lobe-filling supergiant companion star \citep{Reynolds1993, vanderMeer2007}.
The pulsar has a spin period of approximately $0.7$ s (frequency $\nu = 1.4$ Hz) and a variable spin-up rate of around $2.5 \times 10^{-11}$~Hz~s$^{-1}$ \citep{Lucke1976, Inam2010, HuMS2019}.
SMC X-1 is an eclipsing binary with an orbital period of $3.89$ days, which is decreasing over time at a rate of $\dot{P}_{\textrm{orb}} \approx 3.8 \times 10^{-8}$ \citep{Wojdowski1998, Falanga2015, HuMS2019}.
This source also displays quasi-periodic superorbital modulation in the X-ray band, with the modulation period varying between roughly 40 and 65 days \citep[e.g.,][]{Gruber1984, ClarksonCC2003a}.
Notably, ``excursion'' events, during which the superorbital modulation period evolves to approximately 40 days for one to two years, have been observed at least four times \citep{ClarksonCC2003a, Trowbridge2007, Hu2011, DageCC2018, HuMS2019, HuDC2023}.

It has long been suggested that the superorbital modulation of SMC X-1, similar to a couple of X-ray binaries like Her X-1 and LMC X-4 \citep[see, e.g.,][]{ClarksonCC2003b, Charles2008}, are caused by the obscuring of the central compact object by the \rev{radiation-driven} warped or tilted accretion disk \citep{Pringle1996, OgilvieD2001}. 
As long as the accretion disk undergoes precession, its warped regions can periodically obscure the line of sight to the neutron star (NS), with the modulation period corresponding to the precession period of the disk.
Differences in pulse profile shapes and changes in the relative phases between soft X-rays (below 2 keV) and hard X-rays (above 2 keV) support the presence of a warped inner disk structure \rev{in SMC X-1} \citep{Hickox2005, Neilsen2004, BrumbackHF2020, BrumbackVC2023}.
Possible spectral hardening during superorbital transition phases has also been observed, which may be explained by increased absorption near the warped regions of the disk \citep{Trowbridge2007, Hu2011, HuMS2019}.
Recent spectral analysis with \nicer\ reveals a high partial covering fraction, further supporting the scenario in which superorbital modulation arises from substantial obscuration \citep{KaramDT2025}.

In addition, the high luminosity of SMC X-1, approximately $5 \times 10^{38}$ to $1.3 \times 10^{39}$ erg s$^{-1}$, makes it a valuable local analogue of extragalactic ultraluminous X-ray (ULX) pulsars \citep{PradhanMP2020}.
Several ULX pulsars, including M82 X-2 \citep{KongHL2016, BrightmanHB2019}, NGC 5907 ULX-1 \citep{WaltonFB2016, FurstWI2023}, and M51 ULX-7 \citep{VasilopoulosLK2020}, also exhibit superorbital modulations similar to those observed in SMC X-1.
However, in ULX pulsars, these modulations are often attributed to the variations in the mass accretion rate.
In the most extreme case, ULX pulsars may enter the so-called ``propeller'' regime during their superorbital low states, where accretion is inhibited by a centrifugal barrier \citep{IllarionovS1975}.
If this occurs, the steady accretion torque on the NS would cease, leading to a significantly reduced spin-up rate or even a considerable spin-down rate.
While a clear correlation between spin behavior and superorbital phase has not been established in most systems, NGC 5907 ULX-1 did undergo an extended low-flux state lasting three years, during which the pulsar exhibited significant spin-down that can be interpreted as the propeller effect \citep{FurstWI2023}.
However, the 78-day superorbital modulation persists unchanged following the low-flux state, suggesting that regular superorbital modulation is not primarily driven by variations in the mass accretion rate, although intermittent off-states may still occur.
\rev{Another example is NGC 300 ULX-1, a transient pulsating ULX that entered an extended low state in 2019, during which it continued to spin up \citep{VasilopoulosPK2019}. This suggests that the mass accretion rate remained stable despite the drop in observed flux. }

At the same time, the radiation-driven disk warping model \citep{Pringle1996, OgilvieD2001} for superorbital modulation in SMC X-1 was also challenged with a few observational results.
For example, \citet{PradhanMP2020} claimed the detection of the pulsation during the low state, along with variations in the power-law spectral shape and normalization correlated with flux.
This suggests that the flux changes may be, at least in part, intrinsic rather than solely due to obscuration or absorption.
Moreover, the observed increase in low-state flux and the lack of variability in high-state flux during superorbital excursions are not fully consistent with the radiation-driven warped disk model \citep{HuMS2019}.
This does not rule out the disk warping, but suggests that the warp inclination may have no connection to the radiation strength.

To probe whether the mass accretion rate varies across the superorbital cycle, the spin period evolution serves as a valuable proxy.
SMC X-1 exhibits a long-term spin-up trend, indicating a persistent accretion torque acting on the NS \citep{Inam2010, HuMS2019}.
However, detailed measurements of $\dot{\nu}$ within a single superorbital cycle have not yet been performed.
Such analysis is challenging due to the difficulty of obtaining long-term, high-cadence observations from large X-ray observatories. 
All-sky monitors like MAXI are suitable for studying long-term spin frequency evolution \citep[e.g.,][]{HuMS2019}, but their short exposures per orbit (typically a few tens to hundreds of seconds) limit their ability to trace evolution within a single superorbital cycle.
With the increasing feasibility of CubeSat technology, such observational campaigns have become practical.
In this study, we utilize NinjaSat, a 6U CubeSat, to monitor the spin period evolution, flux variability, and spectral behavior of SMC X-1 during a superorbital high state.

\section{Instrument Design and Observation}
\subsection{NinjaSat Observation and Data Reduction}\label{sec:ninjasat_reduction}
NinjaSat is a non-imaging 6U CubeSat X-ray observatory with dimensions of $112.7 \times 237.1 \times 340.5$~mm and a mass of 8 kg \citep{TamagawaEK2025}. 
Launched in 2023, it provides continuous observations of bright X-ray sources and rapid response capabilities for transient follow-up.
Equipped with two Gas Multiplier Counters (GMCs), NinjaSat offers a combined effective area of approximately 32 cm$^2$ at 6 keV, covering the 2--50 keV energy range with a temporal resolution of 61 $\mu$s.
This sub-millisecond timing accuracy has been verified through observations of the Crab pulsar, by comparing the X-ray pulse profile with the radio ephemeris provided by the Jodrell Bank Observatory \citep{LyneJG2015, TamagawaEK2025, OtaTT2025}.
A star tracker guided three-axis attitude control system enables fast and accurate pointing ($<0.1^{\circ}$), while onboard radiation-belt monitors ensure the safe operation of high-voltage components.
This compact but powerful design allows NinjaSat to fill the observational gap between all-sky monitors and large X-ray observatories, enabling high-cadence, high time-resolution data for both time- and frequency-domain studies.

We carried out two observation campaigns using \ninjasat.
The first run took place from July 13 to August 11, 2024, as a test observation. 
It did not cover a complete high state of a superorbital cycle, and several large data gaps severely limited our ability to track the spin frequency evolution in detail.
To address this, a second campaign with a total \textbf{net} exposure of 331 ks \rev{across 485 pointings} was carried out from December 13, 2024, to January 23, 2025, aimed at capturing the full superorbital ascending-high-descending phases.
Although a few data gaps remained, the overall data quality was sufficient to study both spectral variability and spin frequency evolution across the superorbital cycle.

Throughout the entire campaign, only one GMC (GMC1) was in operation.
Data reduction was performed with HEASoft version 6.33.1.
Photon arrival times were corrected to the barycenter of the solar system using the \texttt{barycen} tool with the DE-405 ephemeris.
To verify detector energy calibration, NinjaSat regularly observes  the Crab Nebula and several blank-sky fields.
For the analysis of SMC X-1, we used background data from a blank-sky field at (R.A., Decl.) = ($138.00^\circ$, $15.00^\circ$), known as BKGD\_RXTE3~\citep{JahodaMR2006, RemillardLS2022}, observed between  November 15, 2024 and November 27, 2024, prior to the second monitoring campaign of SMC X-1.
This 12-day observation provides the most statistically robust background data in the vicinity of the SMC X-1 monitoring period, with a total exposure of 114~ks.

The GMC has two readout pads: a circular inner electrode and an annular outer electrode.
Following the event-selection criterion of \citet{TakedaTE2005}, we extracted cleaned events from the inner electrode to ensure a higher signal-to-noise ratio in the 2--20~keV energy range, further dividing them into soft (2--10~keV) and hard (10--20 keV) bands for analysis.
We then corrected the X-ray reduction rate due to the passive collimator of the GMC, taking into account slight misalignments between its optical axis and the source direction, as well as attitude fluctuations during each observation.
The average correction factor over the observation campaign was 1.14, given the collimator's field of view (FoV) of 2.1$^\circ$ at full width at half maximum.
Furthermore, for spectral analysis, we corrected the detector response of the GMC1 following \citet{AoyamaET2025}, whose observations were carried out shortly before the current SMC X-1 and background observations. 
\rev{A uniform correction factor of 0.83 was applied to the effective area of GMC1 across the entire energy range, in addition to the collimator correction described above.}

\subsection{All-Sky Monitoring Data}
To ensure consistency between the \ninjasat\ results and the long-term spin period evolution, and to schedule the \ninjasat\ observations, we used both the Burst Alert Telescope (BAT) onboard the Neil Gehrels Swift Telescope (\swift) and Monitor of all-sky X-ray image (MAXI) data to track long-term flux variability.
In addition, the long-term spin period evolution was derived from MAXI source event data.

\subsubsection{MAXI}
MAXI is a Japanese Experimental Module aboard the International Space Station, equipped with two cameras: the Solid-state Slit Camera (SSC), with a collecting area of 200~cm$^2$ covering the energy range of 0.5 -- 12 keV, and the Gas Slit Camera (GSC), with a collecting area of 5350~cm$^2$ covering 2--30 keV \citep{MatsuokaKU2009}.
The monitoring light curve of SMC X-1\footnote{\url{http://maxi.riken.jp/star_data/J0117-734/J0117-734.html}}, obtained from the GSC and provided by RIKEN, JAXA, and the MAXI team, was extracted in the 2--20 keV energy band.
The spectral hardness information was also available, as the light curves are divided into 2--4, 4--10, and 10--20~keV bands.
We used the one-day binned light curve to calibrate the flux scaling between \swift/BAT and MAXI/GSC, and employed the 6-hour binned light curve for subsequent analysis.

In addition to the all-sky monitoring data, we extracted source events following the procedure described in \citet{HuMS2019}.
Photon events collected by the GSC, which provides high time resolution of 50~$\mu$s \citep{MiharaNS2011}, were used for pulsation searches.
We used 2--20~keV photons collected by GSC units 0, 1, 2, 4, 5, 7, and 8 because other GSCs were out of function \citep{MiharaNS2011, SugizakiMS2011}.
In this work, we present the pulse frequency evolution from MJD 60320 to MJD 60730, covering nine superorbital cycles.
The source photons were extracted from a circular region with a radius of 1$^\circ$ centered on the SMC X-1, ensuring the selection of $\gtrsim$90\% of the source photons \citep{MiharaNS2011}.

\subsubsection{Swift BAT}
The BAT onboard \swift\ is designed primarily to detect gamma-ray bursts, with a large collecting area of 5200~cm$^2$. 
Since 2004, it has also been used to monitor known X-ray sources in the hard X-ray band (15--150~keV) by scanning the entire sky approximately every 96 minutes \citep{BarthelmyBC2005}.
In this study, we used the 15--50~keV light curve provided by the hard X-ray transient monitor program\footnote{\url{https://swift.gsfc.nasa.gov/results/transients/SMCX-1/}} \citep{KrimmHC2013}.

Since MJD 59170, MAXI observations have intermittently experienced interruptions due to occultations by the SpaceX Crew Dragon. 
We used the Swift/BAT light curve to fill these observational gaps.
To align the datasets, we derive a linear scaling factor between the one-day binned MAXI/GSC and \swift/BAT light curves, and convert the BAT count rates into equivalent MAXI count rates.

\section{Results}

\subsection{\ninjasat\ Light Curve of 2024 December Campaign}
We first generated the \ninjasat\ light curve using 1-s time bins and applied collimator corrections as described in Section \ref{sec:ninjasat_reduction}.
Then, the light curve was re-binned over each Good Time Interval (GTI), with exposure ranging from approximately 50 to 1000 seconds.
To guarantee a sufficient signal-to-noise ratio, we excluded GTIs with effective exposure times shorter than 200 seconds.

\begin{figure*}
    \centering
    \includegraphics[width=0.8\linewidth]{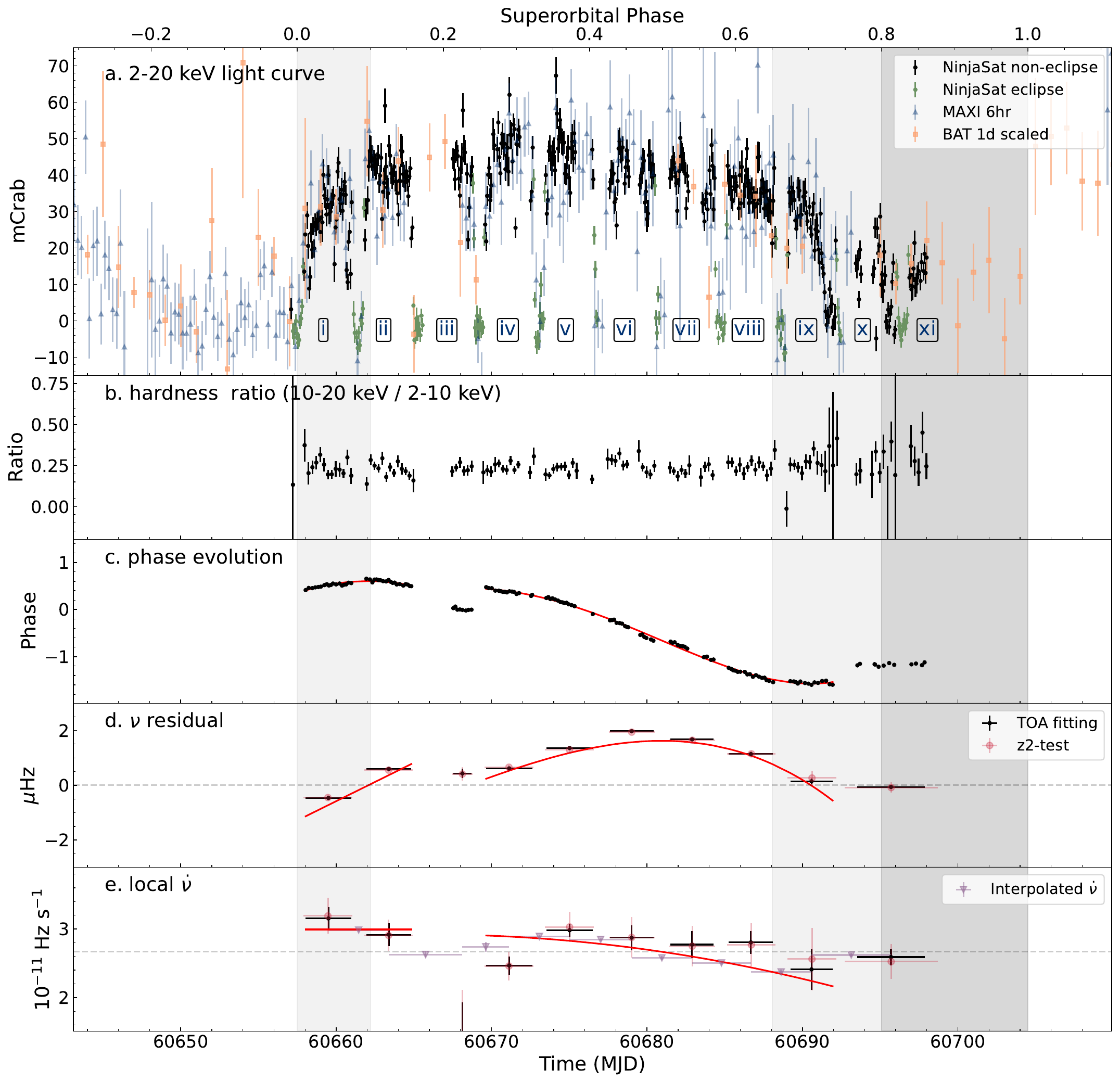}
    \caption{\textbf{a}. Light curve of SMC X-1 obtained with \ninjasat\ (black dots), MAXI (green triangles), and \swift/BAT (orange squares). \ninjasat\ data collected during orbital eclipses are shown in green. The light gray shaded regions indicate the superorbital ascending (MJD 60657--60662) and descending (MJD 60688--60695) states, while the dark gray region marks the superorbital low state (MJD 60695--60704). Eleven observed orbital cycles are labeled {\romannumeral 1} to {\romannumeral 11}. \textbf{b}. Hardness ratio (10--20 keV/2--10 keV) derived from \ninjasat\ observations. \textbf{c}. Evolution of pulse arrival phases relative to a quadratic ephemeris with $\dot{\nu} = 2.668 \times 10^{-11}$ Hz s$^{-1}$, based on the average spin-up rate during this superorbital cycle. \rev{Red curves denote the best-fit quadratic (cycles {\romannumeral 1} to {\romannumeral 2}) and fourth-order polynomials (cycles {\romannumeral 4} to {\romannumeral 9}).} \textbf{d}. Spin period deviations relative to the quadratic model. Spin periods for each orbital cycle are derived from TOA fitting (black points) and cross-validated using the $Z_2^2$-test (orange diamonds). \textbf{e}. Evolution of the local $\dot{\nu}$. \rev{Red curves in panels \textbf{d} and \textbf{e} are derived from timing solutions in panel \textbf{c}.}}
    \label{fig:toa_december}
\end{figure*}

The resulting 2--20 keV light curve is shown in Figure \ref{fig:toa_december}a, with the count rate converted to units of mCrab.
This conversion was performed by dividing the background-subtracted count rates (0.247 counts s$^{-1}$ for the 2--10 keV background and 0.148 counts s$^{-1}$ for the 10--20 keV background) by the corresponding Crab count rates (10.29 counts s$^{-1}$ for 2--10 keV and 1.039 counts s$^{-1}$ for 10--20 keV).
For comparison, the 6-hour binned MAXI light curve and the scaled 1-day binned \swift/BAT light curve are also shown.
In the subsequent timing analysis, we excluded data collected during orbital eclipses (highlighted in green in Figure \ref{fig:toa_december}a), \rev{defined as orbital phase between 0.9 and 1.1 based on the orbital ephemeris from \citealt{HuMS2019}}.

Figure \ref{fig:toa_december} also highlights the superorbital transition states: ascending, defined as $\phi_{\textrm{sup}} = 0.0$--0.1, and descending, defined as $\phi_{\textrm{sup}} = 0.65$--$0.8$, are indicated with a light gray background. 
The superorbital low state is marked with a dark gray background.
A total of eleven orbital cycles were observed by \ninjasat, where the out-of-eclipse intervals are labeled as {\romannumeral 1} through {\romannumeral 11}.

The hardness ratio, defined as the 10--20 keV flux divided by the 2--10 keV flux, for non-eclipse GTIs is shown in Figure \ref{fig:toa_december}b.
The ratio remains largely stable, even during the superorbital ascending and descending states, except for pre-eclipse dips observed during the descending state (around MJD 60692 and MJD 60696).
This suggests that absorption is unlikely to be the dominant factor in superorbital flux variability.

The \ninjasat\ observation covered 41 days, from MJD 60657 to MJD 60698.
Although it did not cover the full superorbital cycle, it captured the complete ascending-high-descending sequence and part of the low state, corresponding to a superorbital phase range of $\phi_{\textrm{sup}} = 0$--0.87.
For this campaign, the start of the superorbital cycle is defined as MJD 60657.5, marking the onset of the rising edge.
The end of the cycle was not captured by either \ninjasat\ or MAXI, but is evident in the \swift/BAT light curve, occurring at MJD 60704.5.

This results in a superorbital cycle duration of 47 days, which is clearly shorter than the average cycle length during non-excursion epochs (approximately 53--57 days), but close to those observed during excursion epochs \citep[typically 45--46 days, see][]{HuMS2019, HuDC2023}.
Thus, this cycle likely represents a mini-excursion, a phenomenon occasionally observed during regular epochs and lasting for one to four cycles.

\subsection{\ninjasat\ Pulsar Timing}
Our goal is to perform phase-connected (coherent) or semi-coherent time-of-arrival (TOA) analysis to track the spin period evolution in detail.
As a preliminary step, we collected photons within each binary orbital period and performed a two-dimensional $Z_2^2$-test to search for both the spin frequency $\nu$ and its derivative $\dot{\nu}$ \citep{BuccheriBB1983}.

To correct for orbital Doppler effects, we used a binary orbital ephemeris with a mid-eclipse time of $T_{\textrm{mid-eclipse}} = $ MJD 60505.7361918, an orbital frequency of $f_{\textrm{orb}} = 0.25696017$~d$^{-1}$, and a frequency derivative of $\dot{f}_{\textrm{orb}} = 2.379 \times 10^{-9}$~d$^{-2}$.
The $f_{\textrm{orb}}$ and $\dot{f}_{\textrm{orb}}$ are based on \citet{HuMS2019}, with the mid-eclipse time shifted to match the epoch of the \ninjasat\ observation.
Additionally, we found that TOAs within a single orbital cycle exhibited residual modulation on the timescale of the orbital period, which was reduced by shifting $T_{\textrm{mid-eclipse}}$ by 100 seconds.
This adjustment allows the ephemeris to effectively correct for orbital Doppler effects across the \ninjasat\ observation window.

After determining the local $\nu$ and $\dot{\nu}$, we divided the binary orbit into multiple segments.
Each segment contains 2--6 \ninjasat\ orbits, and collected approximately 2000--3000 photons.
We then calculated the TOA for each segment using the unbinned maximum-likelihood method \citep{LivingstoneRC2009}.
The phase of each TOA was computed using a long-term average spin frequency of $\nu = 1.4362513237$ Hz and a spin-up rate of $\dot{\nu} = 2.6676 \times 10^{-11}$ Hz s$^{-1}$ \rev{obtained from the \ninjasat\ campaign.}
The resulting TOA phase evolution is shown in Figure \ref{fig:toa_december}c.

To estimate the local values of $\nu$ and $\dot{\nu}$, we fit the TOA phases with a second-order polynomial and present the results in Figures \ref{fig:toa_december}d and \ref{fig:toa_december}e, respectively.
For comparison, spin parameters derived from the local $Z_2^2$-test are also plotted to verify consistency.
Finally, we also estimated the average spin-up rate by calculating the change in $\nu$ between consecutive measurements divided by the elapsed time.
These interpolated $\dot{\nu}$ values are also displayed in Figure \ref{fig:long_term_frequency}d.

From the TOA phase evolution, we identified a possible phase discontinuity around MJD 60670, where the phase exhibits a jump of approximately 0.4 cycles following the orbital eclipse of cycle {\romannumeral 3}.
Unfortunately, the first half of orbital cycle {\romannumeral 3} was not covered by \ninjasat, making it difficult to constrain $\dot{\nu}$ precisely across this interval.
To account for the discontinuity, we divided the TOA phase evolution into two segments: MJD 60658--60662 and MJD 60670--60692, and fit each segment with high-order polynomials (Figure \ref{fig:toa_december}c).
Taking the time derivative of the best-fit phase models, we derived the predicted evolution of $\nu$ and $\dot{\nu}$ (Figures \ref{fig:toa_december}d and \ref{fig:toa_december}e).
These model-predicted trends are consistent with the local $\nu$ and $\dot{\nu}$ values obtained from piecewise TOA fitting and the $Z_2^2$ test results.

Both analyses suggest that the local $\dot{\nu}$ during orbital cycle {\romannumeral 3} appears to drop sharply, although the associated uncertainty is relatively large.
Following this cycle, a significant and abrupt increase in flux is observed, accompanied by a rise in $\dot{\nu}$, suggesting that the spin-up rate may be sensitive to changes in flux.

Another discontinuity is detected during the late descending phase, around MJD 60694, near the end of the orbital cycle {\romannumeral 10}.
This may be attributed either to flux variability or to the contamination of reflection components from the innermost regions of the accretion disk.

\subsection{Pulse Profile Evolution}

\begin{figure}
    \centering
    \includegraphics[width=0.99\linewidth]{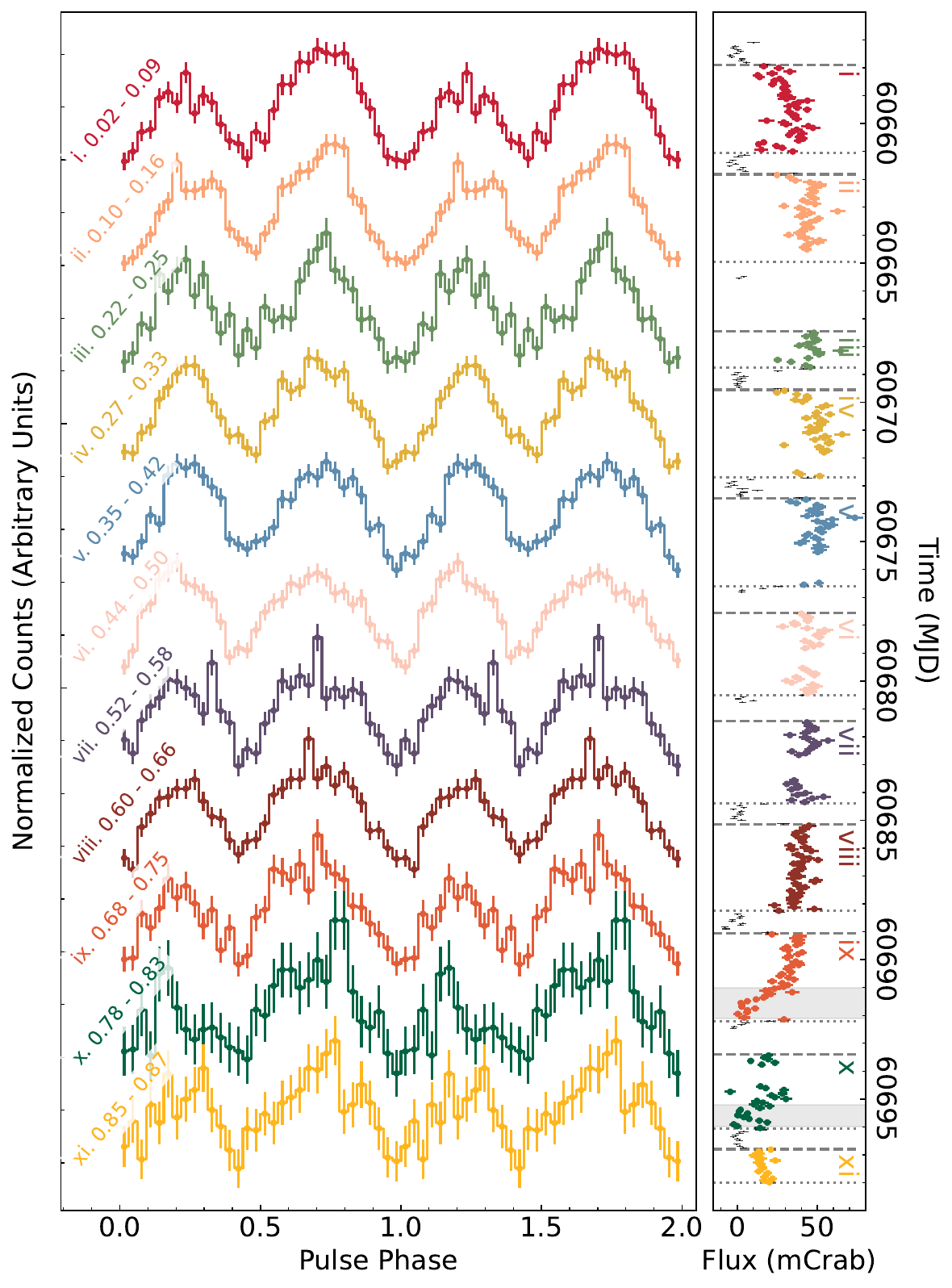}
    \caption{Pulse profile evolution of SMC X-1 observed with \ninjasat. Pulse profiles for 11 orbital cycles ({\romannumeral 1} to {\romannumeral 11}) are shown sequentially and color-coded, with their corresponding superorbital phase ranges labeled. Photon events from orbital eclipses and pre-eclipse dips were excluded from the folded light curves. The right panel displays the \ninjasat\ light curve. Black dots indicate data during orbital eclipses, while out-of-eclipse points are color-coded according to orbital cycle. Dotted lines mark the boundaries of the out-of-eclipse intervals for each orbital cycle. Pre-eclipse dips in cycles {\romannumeral 9} and {\romannumeral 10} are marked by gray-filled regions. Photons collected during these intervals were excluded from the pulse profiles.}
    \label{fig:fold_december}
\end{figure}

To investigate pulse profile variability, we folded the light curve within each orbital cycle using the corresponding local $\nu$ and $\dot{\nu}$ values.
Of the 11 orbits observed with \ninjasat, cycles {\romannumeral 3} and {\romannumeral 11} were only partially covered due to technical issues and scheduling constraints.
Additionally, several orbits, including {\romannumeral 4}, {\romannumeral 5}, {\romannumeral 6}, {\romannumeral 7}, and {\romannumeral 10}, contained significant data gaps.
Nevertheless, we successfully derived the 2--20 keV pulse profiles for each orbit, which are presented in Figure \ref{fig:fold_december}.

All pulse profiles exhibit a double-peaked structure, consisting of a narrower peak (P1) spanning approximately 0.45 cycles and a broader peak (P2) spanning about 0.55 cycles.
At the beginning of the superorbital cycle, P1 has a lower amplitude than P2, but its amplitude gradually increases with superorbital phase.
The two peaks reach similar amplitudes at $\phi_{\rm{sup}} \approx 0.25$--0.35.
Beyond $\phi_{\rm{sup}} \approx 0.35$, P1 becomes more prominent than P2, peaks in relative strength around $\phi_{\rm{sup}} \approx 0.5$, and then begins to decline.
Pulsations remain detectable at $\phi_{\rm{sup}} \approx 0.85$--0.87, and the pulse profile in this phase range resembles that observed at $\phi_{\rm{sup}} \approx 0.02$--0.09, though with greater uncertainties due to limited photon statistics.

\begin{figure}
    \centering
    \includegraphics[width=0.9\linewidth]{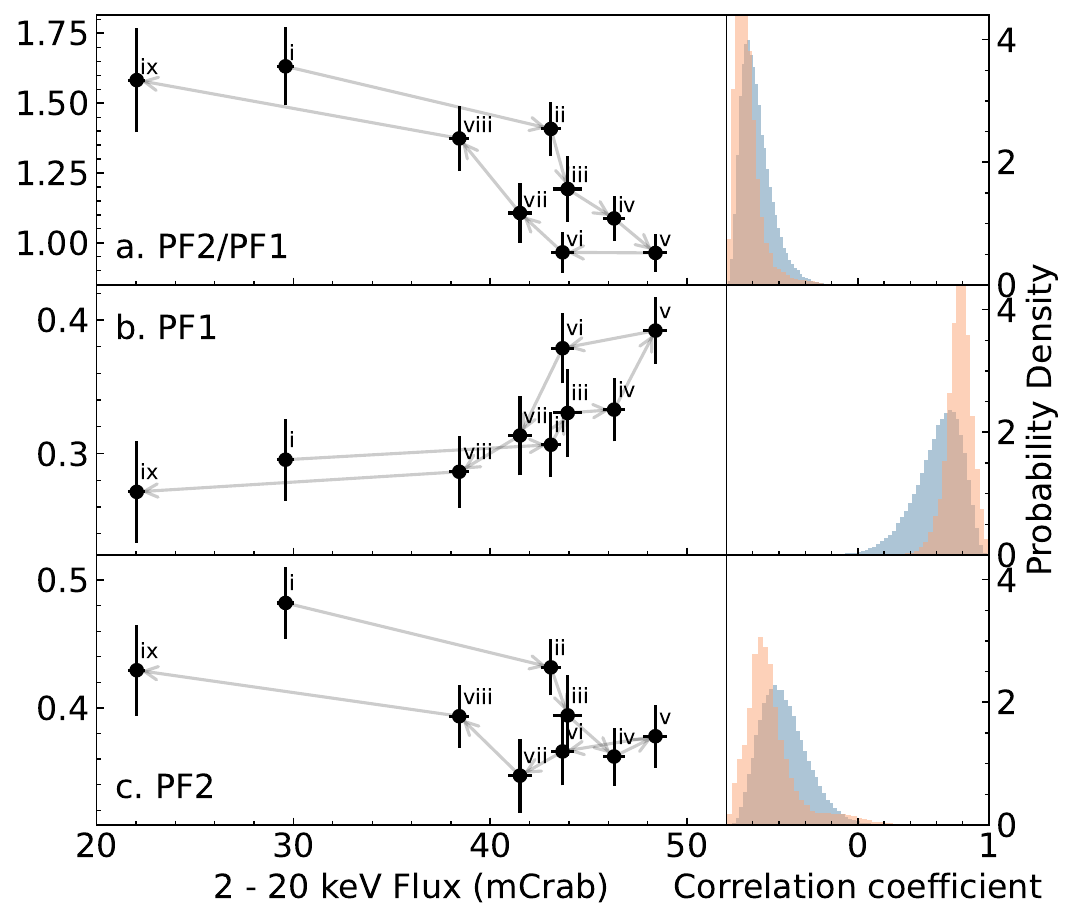}
    \caption{Flux dependence of \textbf{a}. peak ratio, \textbf{b}. pulse fraction of P1, and \textbf{c}. pulse fraction of P2. Gray arrows indicate the evolutionary track of each parameter from orbital cycle {\romannumeral 1} through {\romannumeral 9}. Right panels show probability density distributions from Monte Carlo simulations (blue) and bootstrapped samples (orange).}
    \label{fig:pf_correlation}
\end{figure}

This pulse profile behavior is consistent with the findings of \citet{PradhanMP2020}, and the continuous monitoring by \ninjasat\ enables us to trace its evolution within a single superorbital cycle.
To parameterize the pulse profile, we define the pulse fraction (PF) of P1 as $\rm{PF1} = (P_{\rm{max, 1}} - P_{\rm{UP}}) / (P_{\rm{max}} + P_{\rm{UP}})$, where $P_{\rm{max, 1}}$ is the averaged count rate \rev{over a 0.15-cycle interval centered on} the peak of P1, $P_{\rm{UP}}$ is the mean count rate {over $\phi_{\rm{spin}}=0.95$--1.05 and 0.4--0.5}, and $P_{\rm{max}}$ is the peak count rate of the entire spin cycle, \rev{which equals $P_{\rm{max, 1}}$ if P1 is stronger than P2, or $P_{\rm{max, 2}}$ otherwise}. 
Similarly, the PF of P2 is given by $\rm{PF2} = (P_{\rm{max, 2}} - P_{\rm{UP}}) / (P_{\rm{max, 2}} + P_{\rm{UP}})$. 
To characterize the relative strength of these two peaks, we define the peak ratio as $\textrm{peak ratio} = \rm{PF2} / \rm{PF1} = (P_{\rm{max, 2}} - P_{\rm{UP}}) / (P_{\rm{max, 1}} - P_{\rm{UP}})$.
Due to limited photon statistics, the parameters in orbital cycles {\romannumeral 10} and {\romannumeral 11} were poorly constrained.
Therefore, we carried out correlation analysis on orbital cycles {\romannumeral 1} to {\romannumeral 9} only. 

\begin{figure*}
    \centering
    \includegraphics[width=0.8\linewidth]{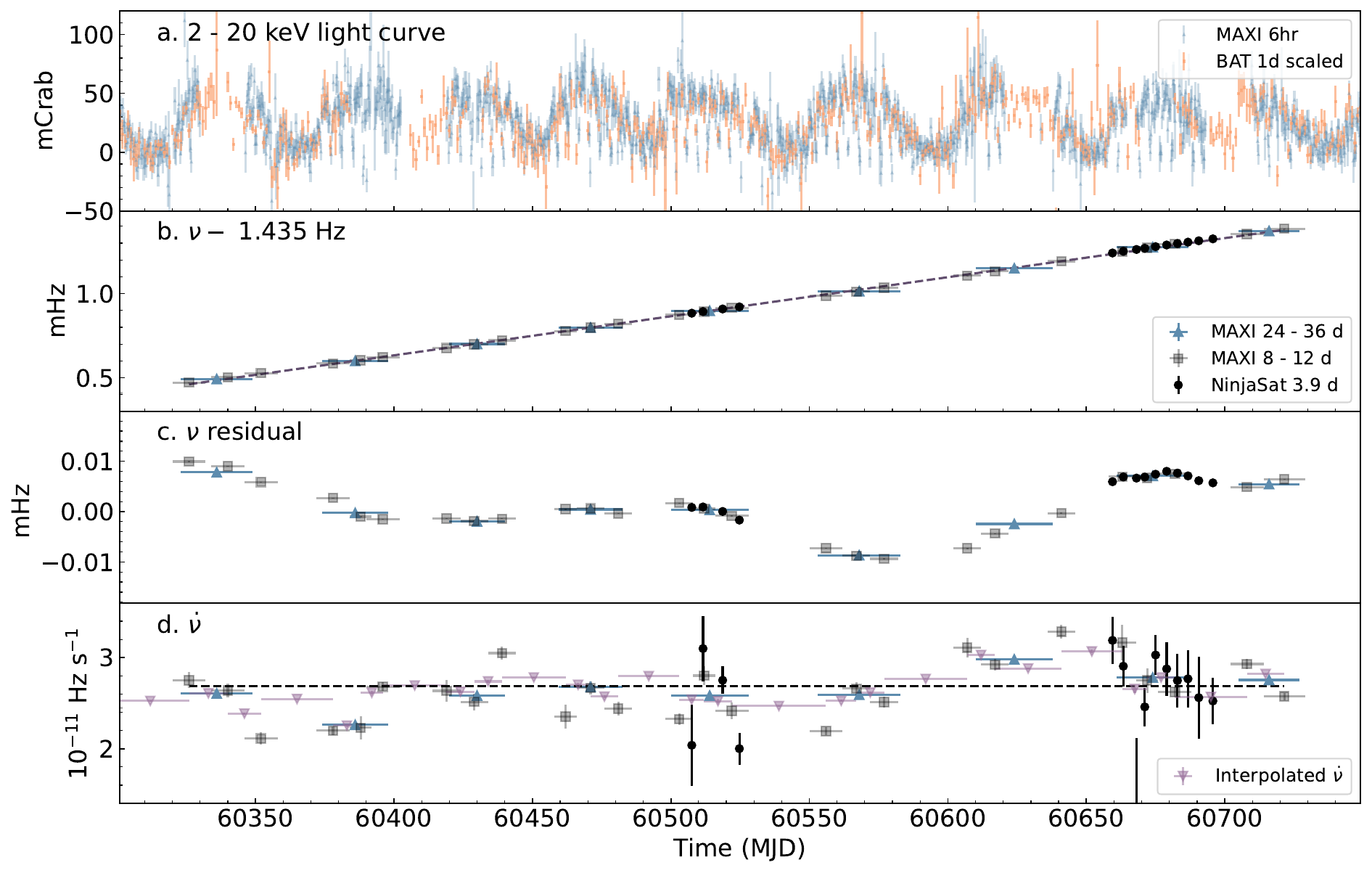}
    \caption{Long-term spin evolution of SMC X-1 observed with \ninjasat\ and MAXI. \textbf{a}. The 2--20 keV MAXI 6-hr binned light curve is plotted in blue, and the scaled 1-d binned \swift/BAT light curve is plotted in orange. \textbf{b}. Spin frequency ($\nu$) evolution. Blue triangles represent values averaged over individual superorbital high state using MAXI data; gray squares show values computed over 8--12 day intervals; black dots indicate \ninjasat\ measurements for each binary orbital cycle. The black dashed line is a linear fit to the 8--12 day MAXI $\nu$ measurements. \textbf{c}. Residuals of $\nu$ after subtracting the best-fit linear trend shown in panel \textbf{b}. \textbf{d}. Evolution of the spin-up rate ($\dot{\nu}$). The dashed line represents the best-fit average $\dot{\nu}$ across the nine superorbital cycles, as determined from the linear model in panel \textbf{b}. }
    \label{fig:long_term_frequency}
\end{figure*}

Figure \ref{fig:pf_correlation} shows the flux dependence of PF1, PF2, and the peak ratio.
The peak ratio ranged from approximately 0.9 to 1.6 during the \ninjasat\ campaign.
Meanwhile, PF1 showed slightly lower values (0.3--0.45) compared to PF2 (0.35--0.5)
A negative correlation is found between flux and peak ratio, with a Pearson coefficient of \rev{$\rho=-0.84$ and a p-value of $p=0.004$}, corresponding to a $\approx3\sigma$ significance level. 
The correlation between flux and PF1 is marginally positive \rev{($\rho = 0.75$, $p = 0.02$)}, whereas flux and PF2 show a relatively weaker anti-correlation \rev{($\rho = -0.66$, $p = 0.05$)}.

To further evaluate the influence of uncertainties and potential outliers, we performed both Monte Carlo and bootstrap simulations.
For the Monte Carlo simulations, we generate simulated samples by assuming the uncertainties in the flux and peak ratio as the diagonal elements of a bivariate Gaussian covariance matrix. 
For the bootstrap analysis, we randomly resample the data set with replacement, keeping the sample size constant.
After $10^5$ simulations for each method, the resulting distributions of $\rho$ are shown in the right panel of Figure \ref{fig:pf_correlation}.

In the case of flux versus peak ratio, only 5 Monte Carlo samples and 91 bootstrap samples exhibited positive correlation coefficients, supporting a statistically significant negative correlation.
For the correlation between flux and PF1, 533 Monte Carlo samples and 48 bootstrap samples had $\rho < 0$, suggesting a moderate positive correlation when measurement uncertainty is taken into account.
Finally, in the flux-PF2 correlation, 580 Monte Carlo and 2374 bootstrap samples yielded positive $\rho$, indicating a marginal significant \rev{negative} correlation.
We also tested the impact of energy selection by increasing the lower bound of energy selection, producing pulse profiles in the 3--20 keV and 4--20 keV bands.
The \rev{positive} correlations between PF1 and flux, as well as between the peak ratio and flux, remain prominent.
In contrast, \rev{PF2} shows only a weak positive correlation with flux.
These results imply that the \rev{broader} peak remains relatively stable with flux, or follows a more non-linear relationship than a simple trend.
In contrast, the \rev{narrower} peak shows clear variability linked to the superorbital phase.

We further investigated the mean hardness ratio variability of both pulse peaks and found no significant flux dependences across the superorbital cycle.
For P1, the result is $\rho = 0.45$ ($p = 0.22$), and for P2, it is $\rho = -0.44 (p = 0.24)$, indicating no statistically significant correlation.

\subsection{Long-term Spin Frequency Evolution}
The \ninjasat\ observations suggest that the $\dot{\nu}$ may be flux dependent.
To investigate whether this behavior is part of a longer-term trend, we analyzed MAXI photon event data and searched for $\nu$ and $\dot{\nu}$ over 8--12 day intervals, based on photon statistics.
For comparison, we also calculated average values of $\nu$ and $\dot{\nu}$ for each superorbital cycle.
In addition to MAXI data, we analyzed another \ninjasat\ dataset from a test observation campaign carried out in 2024 July--August, which spanned approximately 20 days.
Similar to the December campaign, we derived $\nu$ and $\dot{\nu}$ for individual orbits, resulting in four additional data points.

The long-term evolution of $\nu$ and $\dot{\nu}$ is shown in Figure \ref{fig:long_term_frequency}.
We present the spin frequency evolution over nine superorbital cycles, covering the interval from MJD 60330 to 60730.
Figure \ref{fig:long_term_frequency}b displays the spin frequency, which exhibits a steady spin-up trend.
Fitting this trend with a linear model yields an average spin-up rate of $\dot{\nu} = 2.69(1) \times 10^{-11}$ Hz s$^{-1}$.
After subtracting this long-term trend, the residuals are shown in Figure \ref{fig:long_term_frequency}c, revealing modulation with a timescale of roughly four superorbital cycles.

Figure \ref{fig:long_term_frequency}d shows the evolution of $\dot{\nu}$, which also displays clear modulation.
Notably, the variations in $\dot{\nu}$ are more abrupt and exhibit greater fluctuations compared to the smoother residuals in $\nu$.
Additionally, the $\dot{\nu}$ modulation seems to lead the $\nu$ residuals by approximately one superorbital cycle.
To fill in data gaps caused by missing data and superorbital low states, and check the consistency of the spin evolution, we also estimated the interpolated spin-up rate from the $\nu$ measured with MAXI.

From the long-term evolution, it is clear that the superorbital cycle observed during the December campaign lies on a plateau in the $\nu$ residual plot, coinciding with a declining trend in the long-term $\dot{\nu}$ across neighboring superorbital cycles.
This suggests that the apparent drop in frequency residual during the ascending and descending phases of the December campaign is likely coincidental, rather than an intrinsic dependence on superorbital phase.

\subsection{Spectral Evolution}
We used \ninjasat\ to perform spectral analysis over this superorbital cycle.
Orbital cycles {\romannumeral 1}, {\romannumeral 2}, {\romannumeral 4}, {\romannumeral 8}, and {\romannumeral 9} were nearly continuously monitored. 
This allows us to divide their non-eclipse phases into two time segments each.
Most importantly, the second half of orbital cycle {\romannumeral 9} likely corresponds to a pre-eclipse dip, enabling a comparison between regular and dip spectra.
Other orbital cycles contain significant data gaps, so they were not subdivided for spectral analysis.
This resulted in a total of 16 time segments used for spectral analysis.

We performed all spectral analyses using XSPEC version 12.14.0.
The absorption model \texttt{tbabs} was adopted, with interstellar abundances set according to \citet{WilmsAM2000}, in which the cross sections was adopted from \citet{VernerYB1993}.
We grouped the spectra to ensure that each spectral bin contained at least five photons, and used Cash statistics for spectral fitting \citep{Cash1979}.

\begin{figure}
    \centering
    \includegraphics[width=0.95\linewidth]{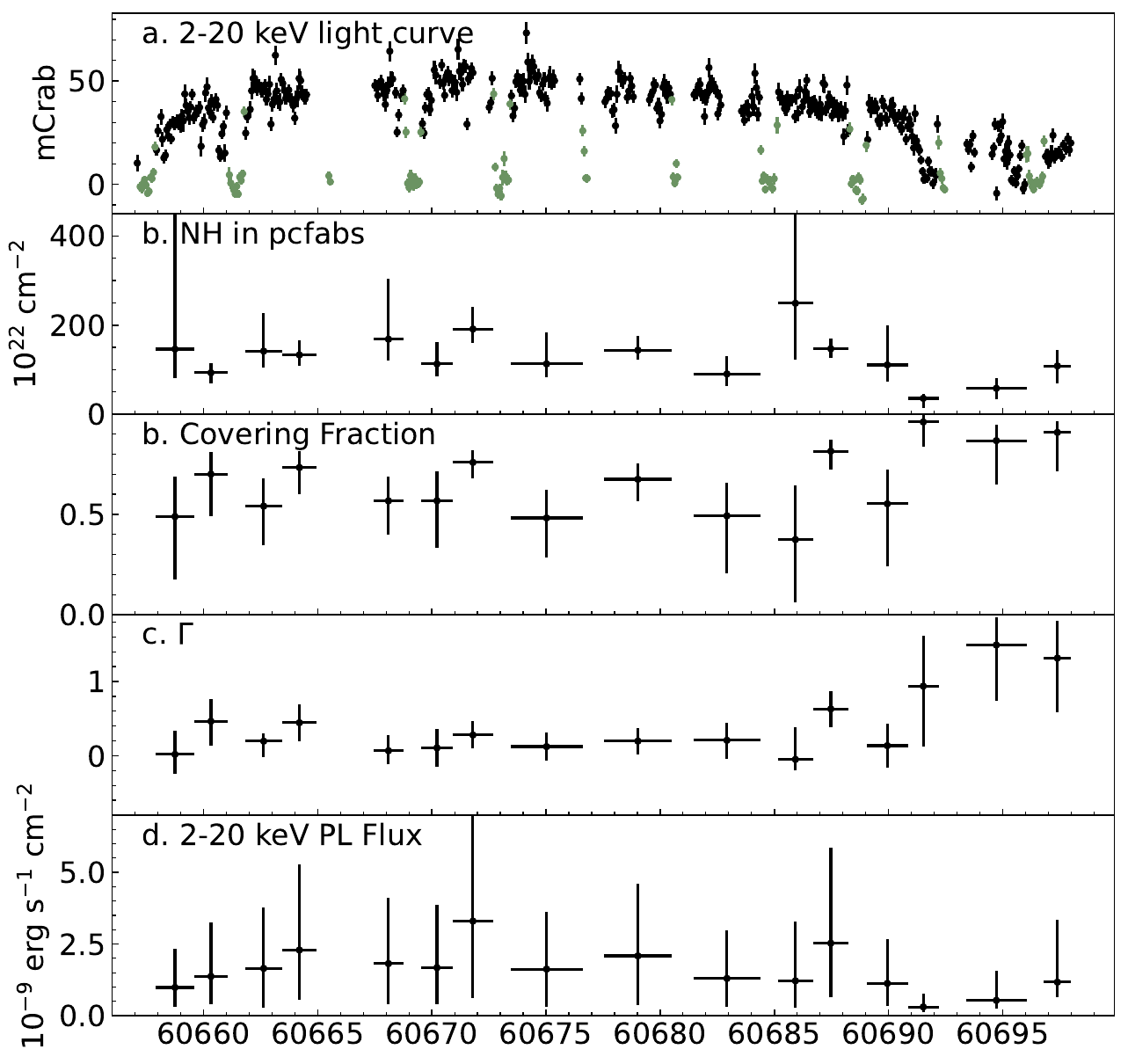}
    \caption{Spectral evolution of SMC X-1 during the \ninjasat\ monitoring campaign from December 2024 to January 2025. Panel a presents the 2--20 keV light curve as a reference. Panels b--d illustrate the corresponding evolution of spectral parameters derived from the best-fit pcfabs*cutoffpl model: b. $N_{\textrm{H}}$ from the pcfabs component, c. partial covering fraction, d. photon index ($\Gamma$), and e. unabsorbed flux (2--20 keV band).}
    \label{fig:spectral_result}
\end{figure}

The X-ray spectrum of SMC X-1 is typically described with a blackbody component, a power-law with a high-energy cutoff, and several emission lines.
The power-law continuum is commonly modeled using cutoffpl \citep[e.g.,][]{PradhanMP2020, KaramDT2025}) or \texttt{highecut*powerlaw} \citep[e.g.,][]{Neilsen2004, Hickox2005}.
Prominent spectral lines often include Fe K$\alpha$, Ne X Ly$\alpha$, Ne IX, and O VIII Ly$\alpha$ \citep{BrumbackHF2020}.
However, these line features are not detectable in \ninjasat\ data due to the limited effective area and the low energy resolution of the gas detector.
Therefore, we tested two continuum models: \texttt{tbabs*pcfabs*cutoffpl} and \texttt{tbabs*pcfabs*highecut*powerlaw}.
In both models, \texttt{pcfabs} is used to represent partial covering by the warped accretion disk.
The blackbody component is not included because its contribution is negligible above 2 keV.

\begin{figure}
    \centering
    \includegraphics[width=0.95\linewidth]{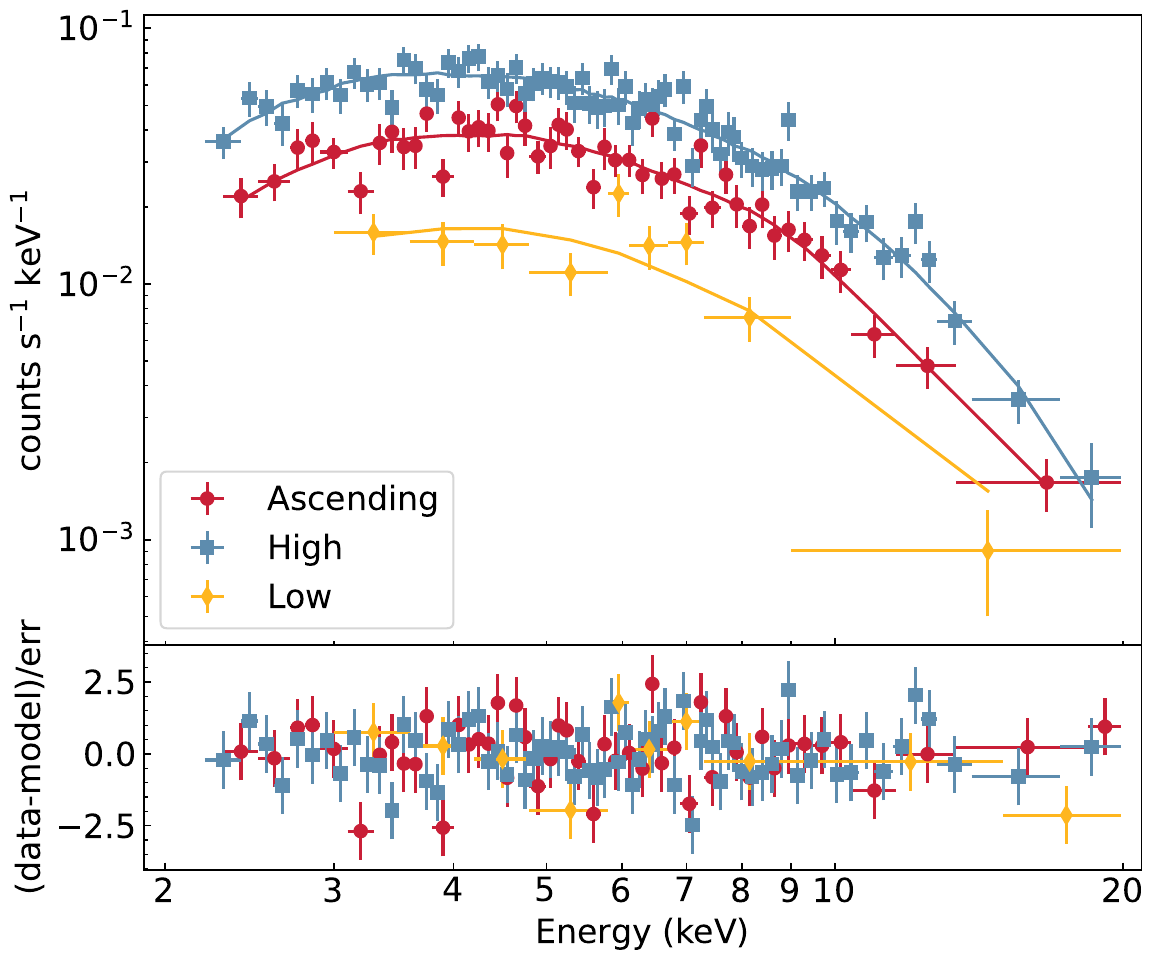}
    \caption{Three example spectra observed with \ninjasat. The spectrum during the ascending state corresponds to the first data point in Figure \ref{fig:spectral_result}; the high state spectrum is taken from the seventh point, and the low state spectrum is from the last data point. The bottom panel displays residuals normalized to data uncertainties. Spectra are rebinned for display to maintain a signal-to-noise ratio above 5 for display purposes. }
    \label{fig:spectra_example}
\end{figure}

We fit all 16 spectra simultaneously.
The hydrogen column density \nh) in the \texttt{tbabs} model was freezed at $2.5 \times 10^{21}$ cm$^{-2}$ \citep{KaramDT2025}, while the $N_{\rm H}$ of the \texttt{pcfabs} component was a free parameter.
The e-folding energy in the \texttt{cutoffpl} model cannot be constrained for individual data sets. 
Therefore, it was linked across all data sets.
This model yielded a total fit statistic of 2713.16 with 2783 degrees of freedom, corresponding to a null hypothesis probability of 0.29.
On the other hand, the \texttt{highecut*powerlaw} model presented difficulty in simultaneously constraining the cutoff energy ($E_c$) and e-folding energy ($E_f$), so we linked these two parameters across all data sets.
This resulted in a total fit statistic of 2717.28 with 2797 degrees of freedom and a higher null hypothesis probability of 0.27.
Both models are statistically acceptable. 
Therefore, we adopt the \texttt{tbabs*pcfabs*cutoffpl} model, with the e-folding energy of exponential rolloff \rev{$E_f = 3.2_{-0.5}^{+0.4}$ keV linked across all data sets}, for reporting the spectral evolution.

The spectral parameter evolution is shown in Figure \ref{fig:spectral_result}.
The column density ($N_{\rm{H}}$) derived from the pcfabs component is generally large \rev{and poorly constrained.}
High covering fractions exceeding 0.9 is found in the superorbital low state 
including two pre-eclipse dips.
The photon index $\Gamma$ and PL flux show no strong dependence on superorbital phase, except during the low states. 
\rev{In addition, the $\Gamma$ is frequently found between 0 and 1, occasionally approaching zero. These are partially consistent with those reported by \citet{DageBN2022} and \citet{KaramDT2025}, though with significantly larger uncertainties. The increased uncertainty may arise from parameter degeneracy between the partial covering $N_{\rm{H}}$ in \texttt{pcfabs} and $\Gamma$, likely due to limited soft X-ray sensitivity ($<$2 keV) and photon statistics of \ninjasat\ for dealing with complex spectral model.} 
Therefore, although a softer spectrum is suggested in the low state, the larger uncertainties make this inconclusive.
These findings indicate that the spectral shape remains broadly stable throughout the superorbital cycle in the \ninjasat\ energy range.

To illustrate the spectral variability of SMC X-1 and demonstrate the observational capabilities of \ninjasat, we present three representative spectra in Figure \ref{fig:spectra_example}.
They represent different superorbital states: the ascending (first half of orbit {\romannumeral 1}), the high state (second half of orbit {\romannumeral 4}), and the low state (first half of orbit {\romannumeral 11}).
They correspond to the first, seventh, and last data points in Figure \ref{fig:spectral_result}.

\begin{figure}
    \centering
    \includegraphics[width=0.95\linewidth]{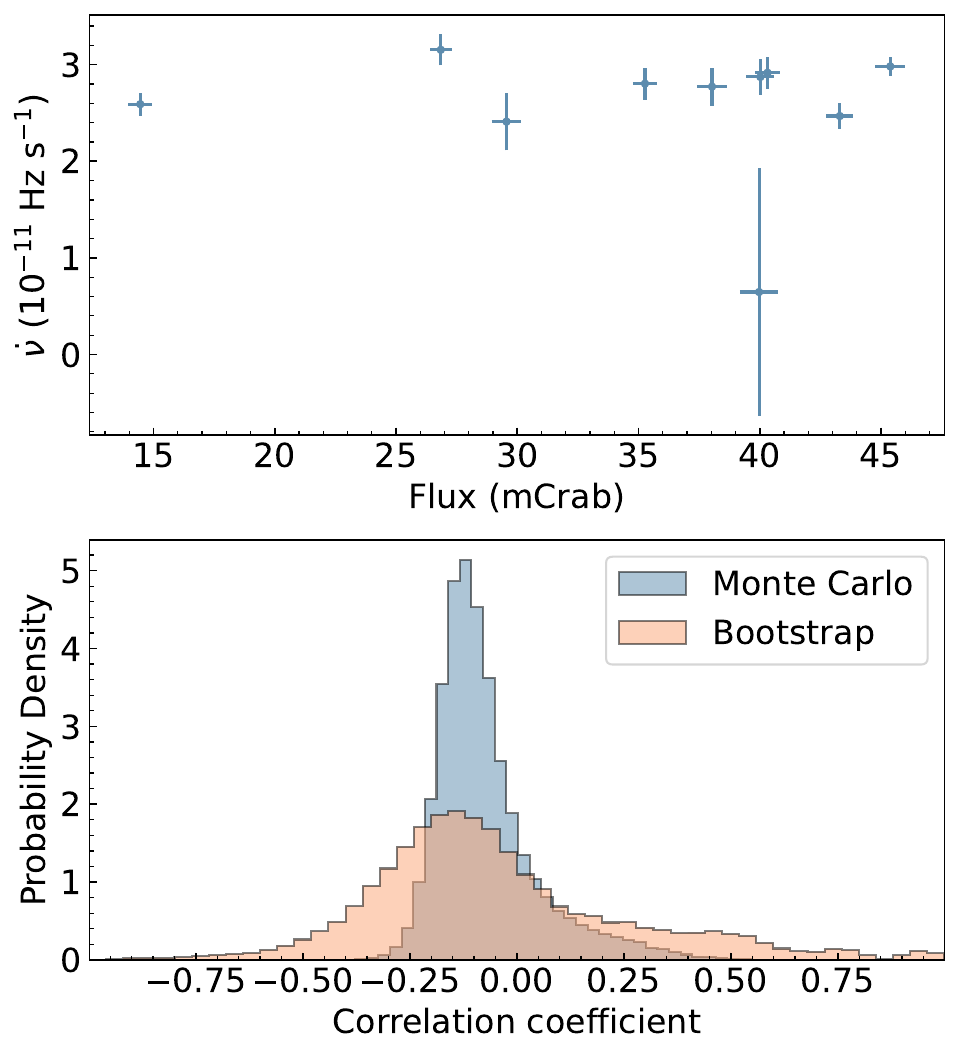}
    \caption{Correlation between $\dot{\nu}$ and 2--20 keV flux. No statistically significant correlation was observed from both Monte Carlo simulations and bootstrap resampling.}
    \label{fig:corr_nudot_flux}
\end{figure}

\section{Discussion}

With the \ninjasat\ observations, we successfully traced the spin frequency evolution of SMC X-1 across a superorbital cycle lasting just 47 days, which indicates a mini excursion.
We found that while the spin-up rate varies, this variability appears to be governed by longer-term trends rather than being directly tied to the superorbital phase.
Despite significant X-ray flux variations between approximately 10 and 50 mCrab, the hardness ratio remains relatively stable, between 0.2 and 0.3, with slight increases observed only during pre-eclipse dips.
The flux during the superorbital low state in the cycle between 60650 and
60700 is approximately 10--20 mCrab, higher than in typical low states during regular epochs.
Pulsations remain detectable even during this low state, and we observed clear evolution in the pulse profile as a function of superorbital phase.
All the spectra can be described with a cutoff power law, where the partial covering fraction increased to be higher than 90\% during the pre-eclipse dips and superorbital low state. 
These observational results provide hints at the nature of superorbital modulation, the interaction between the magnetosphere and the inner accretion disk, and the geometry of the disk itself.
We explore the implications of these findings in the following discussion.

First, although the spin-up rate of SMC X-1 is variable, the spin frequency increases monotonically throughout the observation period.
This indicates that the magnetospheric radius ($r_{\mathrm{m}}$), which is believed to be the innermost radius of the accretion disk, remains well within the corotation radius ($r_{\mathrm{co}}$).
Under this condition, the spin-up rate is expected to be proportional to the mass accretion rate ($\dot{m}$). 
Following \citet{BildstenCC1997}, the relationship can be expressed as:
\begin{equation}
\dot{\nu} \approx 1.6\times10^{-13} \textrm{\,s}^{-2} 
\left(\frac{\dot{m}}{10^{-10}M_{\odot}\textrm{\,yr}^{-1}} \right)
\times\left(\frac{\nu}{Hz} \right)^{-\frac{1}{3}}\left(\frac{r_{\mathrm{m}}}{r_{\mathrm{co}}} \right)^{\frac{1}{2}}. 
\end{equation}
Assuming standard magnetospheric accretion, the luminosity is also proportional to $\dot{m}$, implying that $\dot{\nu}$ should also be proportional to the luminosity.
In our observations, the X-ray flux varies by at least a factor of five.
However, the measured change in spin-up rate between high and low states is only about 13(5)\% if we comparing orbital cycles {\romannumeral 5} and {\romannumeral 10}–{\romannumeral 11}.
This discrepancy suggests that the superorbital modulation is primarily due to geometric effects rather than intrinsic changes in the accretion rate.

This can be further examined through a correlation analysis between the flux and $\dot{\nu}$ (see Figure \ref{fig:corr_nudot_flux}).
The Pearson correlation coefficient ($\rho = -0.1$, p = 0.76), supported by results from Monte Carlo and bootstrap methods, indicates no statistically significant correlation.
The lack of correlation implies that $\dot{\nu}$ is not directly related to observed flux, supporting the interpretation that superorbital modulation is not primarily driven by changes in the mass accretion rate.

In addition, an intriguing phase discontinuity is observed between orbital cycles {\romannumeral 3} and {\romannumeral 5}.
We observed a $\approx10$\% increase in $\dot{\nu}$ across these orbital cycles, confirmed by both direct and interpolated measurements (Figure \ref{fig:toa_december}e). 
This event aligns with a $\approx10$\% increase in X-ray flux (Figure \ref{fig:pf_correlation})
This coincidence suggests a potential short-term link between flux and spin-up rate during the superorbital high state, where the covering fraction is relatively stable.
A similar discontinuity occurs between cycles {\romannumeral 9} and {\romannumeral 11}, but strong flux variations due to superorbital modulation and pre-eclipse dips prevent further interpretation.

Since the superorbital modulation is unlikely to be primarily driven by mass accretion rate change, the warped accretion disk model remains one of the most plausible explanations.
However, this model also faces certain challenges.
One key issue is the lack of significant spectral or hardness ratio variability across the superorbital cycle, except during pre-eclipse dips.
Although previous studies have suggested spectral hardening in the superorbital low state, such hardening has only been observed during the deepest low states and is associated with large uncertainties \citep{Trowbridge2007, HuMS2019}.
Notably, no clear spectral hardening and spectral parameter change are seen during the superorbital transition phases.
This suggests that the warped region of the accretion disk is highly opaque across the entire X-ray bands.
The gradual flux changes observed during the superorbital transitions, as well as the variability within the low state, may instead be attributed to variations in the partial covering fraction.
Recent spectral analyses with \nicer\ have revealed an anti-correlation between observed flux and partial covering fraction, supporting this interpretation \citep{KaramDT2025}.

If the superorbital modulation is indeed caused by partial obscuration from an opaque, warped accretion disk, the inner disk must lie close enough to make the NS to have a non-negligible angular size.
The corotation radius, estimated from the NS spin frequency, is approximately 1300 km. In contrast, the inner disk radius may be much smaller.
For instance, \citet{PaulNE2002} estimated an inner radius of 300--400 km based on modeling of the soft spectral component, though this does not fully explain the detection of pulsations below approximately 2 keV.
If this small inner radius is correct, the NS would have an angular size of roughly 4 degrees as seen from the inner disk. 
This is sufficient for partial covering.
Under this geometry, precession of the warped disk could produce gradual transitions between high states (NS fully visible) and low states (partially or mostly obscured).
Furthermore, small variations in the disk structure could result in a shallower low state if the warp is flatter and fails to fully obscure the NS.
While this scenario is plausible, understanding the link between short superorbital cycle lengths and the observed shallow low states remains crucial for a comprehensive understanding of this system.

Although the superorbital modulation is unlikely to be caused by changes in the mass accretion rate, i.e., not driven by the ``propeller'' effect, this does not rule out the possibility of a supercritical accretion scenario \citep[see][for a review]{KaaretFR2017}.
\rev{In this scenario}, the inner accretion disk becomes geometrically thick due to radiation pressure \citep{King2008}.
If the inner disk is asymmetric to the orbital plane and undergoes precession, it can (quasi-)periodically obscure the X-ray emission from the NS, similar to the mechanism proposed in the radiation-driven warp model.
In this scenario, optically thin material is likely blown away from the upper layer of the accretion disk, leaving behind an opaque structure.
This could naturally explain the lack of significant spectral changes during the superorbital transitions, as the remaining disk materials are optically thick in the X-ray band.
In addition, since the intrinsic luminosity of SMC X-1 is as high as of $1.3 \times 10^{39}$ erg s$^{-1}$, the inner accretion disk is likely to be highly ionized. 
Under these conditions, the observed decrease in flux during the low state can also be explained by nearly energy-independent electron scattering, rather than increased absorption or spectral hardening.
A similar mechanism is also expected in the precessing ring tube model \citep{Inoue2019}.
Finally, the ejected material is expected to form a funnel-shaped wind, driving ultra-fast outflows at semi-relativistic speeds \rev{as seen in NGC 300 ULX-1 \citep{KosecPW2018}}.
Such features may be detectable with future high-spectral-resolution X-ray missions like the X-Ray Imaging and Spectroscopy Mission.

The hard X-ray ($\gtrsim 2$ keV) pulse profile of SMC X-1 is generally characterized as double-peaked and stable across the superorbital phase.
In contrast, the soft X-ray component ($\lesssim$2 keV) shows significant variability.
Their phase-dependent difference has long supported the warped disk model \citep{ Neilsen2004, Hickox2005}.
However, variability in the hard X-ray band has received limited discussion.
Our \ninjasat\ observations confirms the result obtained in \citet{PradhanMP2020} that the hard X-ray pulse profile is modulated by superorbital phase, which is not dominated by intrinsic flux variability.
If the NS is effectively a point source as seen from the inner accretion disk, geometric effects cannot account for the observed variability.
On the other hand, if the NS has a non-negligible angular size, the two pulsar beams may experience differing covering fractions depending on the superorbital phase.
\rev{Such a scenario has been successfully developed to explain the pulse profile variability in Her X-1 \citep{ScottLW2000}. Given that the pulse profile of SMC X-1 is quite different from that of Her X-1 and that no corresponding simulations have been performed, this hypothesis could be tested in future work through detailed theoretical modeling and superorbital/pulse phase-resolved spectroscopy.}

Precession of the NS spin axis presents a plausible alternative explanation for the pulse profile evolution, similar to the mechanism proposed for Her X-1 \citep{StaubertKP2009, StaubertKV2013}.
In the case of Her X-1, precession of the NS is considered a more stable clock than the disk precession, with the two synchronized through a feedback mechanism.  
This coupling may weaken when the inner inclination of the accretion disk decreases.
This scenario not only explains pulse profile evolution within a superorbital cycle but also account for correlations between spin period changes and the appearance of anomalous low states. 
\rev{X-ray polarization observations with the Imaging X-ray Polarimetry Explorer (IXPE) have provided evidence for free precession of the neutron star in Her X-1 \citep{DoroshenkoPT2022, HeylDG2024}.}
A similar behavior is also observed in SMC X-1.
MAXI data reveal a spin-up rate acceleration before the most recent superorbital excursion \citep[Figure 2 in][]{HuDC2023}, supporting an inside-out mechanism where the NS behavior influences disk structure.
However, this connection is not always observed. 
No such change in spin frequency was detected around the shorter 2014 excursion \citep{HuMS2019, HuDC2023}.
Future long-term monitoring will be crucial for testing the consistency between pulse profile evolution and superorbital modulation.
Finally, the hypothesis of neutron star spin-axis precession can also be independently tested using X-ray polarization with IXPE over the superorbital cycle.

\section{Summary}
In this work, we present the first high-cadence, long-term timing and spectral monitoring of SMC X-1 across nearly an entire superorbital cycle using \ninjasat. 
This unique observational campaign allowed us to continuously trace the spin frequency and spin-up rate, and enabled a direct test of whether spin evolution correlates with X-ray flux across different superorbital states. 
Combining \ninjasat\ and MAXI data, we find that the spin-up rate during the high state remains consistent with the long-term average and shows no significant correlation with instantaneous flux, even during low and transitional states. 
This strongly suggests that the superorbital modulation is not driven by intrinsic changes in the mass accretion rate, but is instead geometric in origin, likely caused by obscuration from a precessing, optically thick warped accretion disk or energy-independent electron scattering.

Our spectral analysis supports this interpretation. 
Both the hardness ratio and the overall spectral shape remain stable throughout the superorbital cycle, including the low states. 
This spectral stability implies that observed superorbital flux changes are not accompanied by changes in absorption \rev{of a neutral fully covering absorber}, consistent with the partial covering scenario.

A possible short-term link between $\dot{\nu}$ and flux is observed in the high state, where both increased by approximately 10\%. 
This result suggests that local flux variability could be caused by changes in the mass accretion rate. 
A longer-term trend (longer than a single superorbital cycle) is also observed in the combined \ninjasat\ and MAXI campaign, although further analysis is needed to test the connection between flux and spin-up rate over such timescales.

Finally, we detect superorbital-phase-dependent variations in the 2--20 keV pulse profile.
These changes are primarily associated with variations in the pulse fractions of the two peaks in the profile. 
This behavior could be interpreted by either the variability of the covering fraction of two hotspots, or the precession of the NS spin axis. 
Continued long-term monitoring, detailed theoretical modeling, and polarization observation will be essential to test the stability and origin of this pulse profile evolution.

Beyond these scientific insights, our study demonstrates the capability of CubeSats like \ninjasat\ for long-term, high-cadence observations of bright X-ray sources. 
Future CubeSat missions with better energy resolution, broader spectral coverage, or multi-wavelength capability could greatly expand our understanding of accreting compact objects.

\begin{acknowledgments}
\rev{We thank the anonymous referee for valuable comments that improved this paper.} 
This project was supported by the Japan Society for the Promotion of Science (JSPS) KAKENHI (JP17K18776, JP18H04584, JP20H04743, JP24K00673, JP25KJ0241). 
C.-P.H. acknowledges support from the National Science and Technology Council in Taiwan through grant 112-2112-M-018-004-MY3 and JSPS Invitational Fellowship for Researcher in Japan (ID: S24075). 
T.E. was supported by the “Extreme Natural Phenomena” RIKEN Hakubi project, and the JST Japan grant number JPMJFR202O (Sohatsu). 
N.O. was supported by RIKEN Junior Research Associate Program.
A.A. and S.I. were supported by RIKEN Student Researcher Program.
\end{acknowledgments}

\facilities{\emph{NinjaSat}, \emph{Swift}, MAXI}
\software{HEASOFT}

\bibliographystyle{aasjournal}
\bibliography{reference}

\end{document}